\documentclass[runningheads]{article}

\usepackage{arxiv}

\usepackage[utf8]{inputenc} 
\usepackage[T1]{fontenc}    
\usepackage{hyperref}       
\usepackage{url}            
\usepackage{booktabs}       
\usepackage{amsfonts}       
\usepackage{nicefrac}       
\usepackage{microtype}      
\usepackage{lipsum}
\usepackage{graphicx}
\usepackage[utf8]{inputenc}
\usepackage{caption}
\usepackage{float}
\usepackage{refstyle}
\usepackage[export]{adjustbox}
\usepackage{amsmath}
\usepackage{ulem}
\usepackage{fix-cm}
\usepackage{array}
\usepackage{caption}
\usepackage{subcaption}
\usepackage{longtable}
\usepackage{changepage}
\usepackage{textcomp}
\usepackage{multirow}
\usepackage{fancyhdr}
\usepackage{comment}
\graphicspath{ {./images/} }

\begin{document}

\title{Identification of Cervical Pathology in Colposcopy Images using Adversarially Trained Convolutional Neural Networks}


\author{
 Abhilash Nandy \\
  Indian Institute of Technology Kharagpur\\
  West Bengal\\
  India \\
  \texttt{raj12345@iitkgp.ac.in} \\
  \And
 Rachana Sathish \\
  Indian Institute of Technology Kharagpur\\
  West Bengal\\
  India \\
  \texttt{rachana.satish@iitkgp.ac.in} \\
  \And
 Debdoot Sheet \\
  Indian Institute of Technology Kharagpur\\
  West Bengal\\
  India \\
  \texttt{debdoot@ee.iitkgp.ac.in} \\
}
\maketitle
\begin{abstract}
Various screening and diagnostic methods have led to a large reduction of cervical cancer death rates in developed countries. However, cervical cancer is the leading cause of cancer related deaths in women in India and other low and middle income countries (LMICs) especially among the urban poor and slum dwellers. Several sophisticated techniques such as cytology tests, HPV tests etc. have been widely used for screening of cervical cancer. These tests are inherently time consuming. In this paper, we propose a convolutional autoencoder based framework, having an architecture similar to SegNet which is trained in an adversarial fashion for classifying images of the cervix acquired using a colposcope. We validate performance on the Intel-Mobile ODT cervical image classification dataset. The proposed method outperforms the standard technique of fine-tuning convolutional neural networks pre-trained on ImageNet database with an average accuracy of $73.75\%$

\end{abstract}


\section{Introduction} \label{introduction}

Cervical cancer occurs when the epithelial cells lining up the cervix grow abnormally and invade neighboring tissues and organs of the body. Generally associated with infection of the human Papillomavirus (HPV)~\cite{screening2018cervical}, associated with unhygienic sanitary conditions, cervical cancer accounts for the most common cause of cancer cases and deaths reported among urban poor in low and middle income countries (LMICs)~\cite{bray2018global}. Anatomically, the cervix is constituted of endocervix which is closest to the uterus and the part next to the vagina is the ectocervix. While endocervix is made up of columnar epithelium, the ectocervix is made up of stratified squamous epithelium cells and the boundary region between them called as transformation zone under pathological variation exhibits most of the origin of squamous cell carcinoma or dysplasia. Post its onset, the cancer can turn out to be completely ectocervical, partially ectocervical and partially endocervical, or completely endocervical. Since management of successive diagnostic, treatment and prognostic protocols are different for each of them~\cite{mark2018expert}, identification of their specific type is of immense importance. 

The clinical protocol makes use of an optical imaging device termed colposcope that provides magnified view of the cervix. This device is cost effective with low cost of ownership and commonly found across most primary healthcare centers in LMICs and rest of the world. The challenge however is lack of trained Gynecologists available at these equipped centers to report the screening procedure. It has accordingly been observed that due to this, screening tests report high cases of false negatives~\cite{screening2018cervical}, which leads to late administration of intervention at advanced stages which often is irrecoverable.

The onset of metaplastic changes in the transformation zone when it is ectocervical and fully visible, as in Fig. \ref{fig:cervix_a}, second being 
partially ectocervical and partially endocervical and fully visible, as shown in Fig. \ref{fig:cervix_b}, and the third being completely endocervical and is not fully visible, as shown in Fig. \ref{fig:cervix_c}, are the three major manifestations that require identification.

\begin{figure}[t]
\centering
    \begin{subfigure}[b]{0.32\textwidth}
        \includegraphics[width=\linewidth]{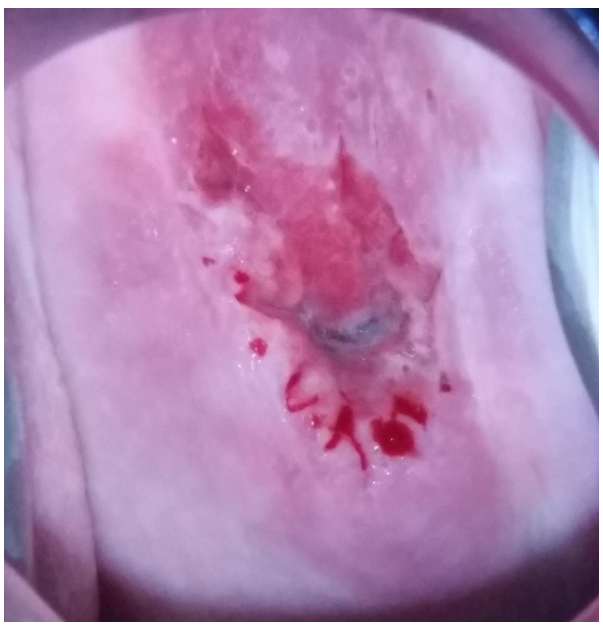}
        \caption{(a) Type 1}
        \label{fig:cervix_a}
    
    \end{subfigure}%
    \begin{subfigure}[b]{0.25\textwidth}
        \includegraphics[width=\linewidth]{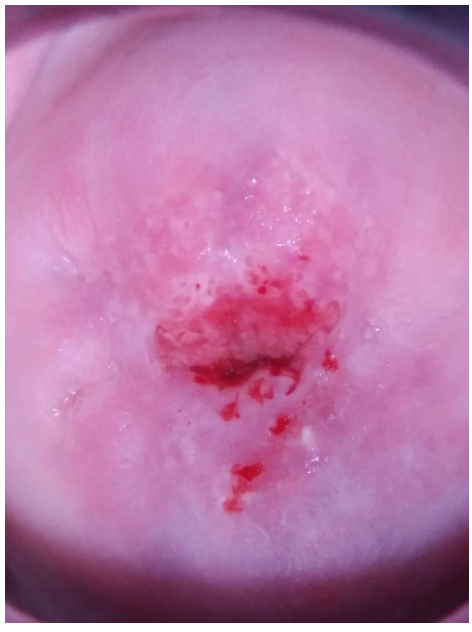}
        \caption{(b) Type 2}
        \label{fig:cervix_b}
    \end{subfigure}%
    \begin{subfigure}[b]{0.25\textwidth}
        \includegraphics[width=\linewidth]{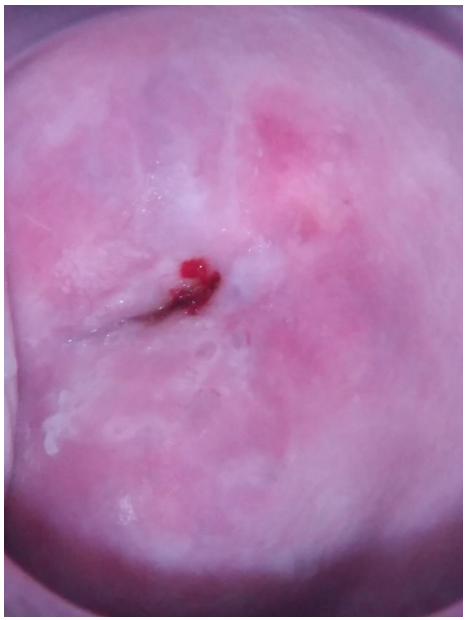}
        \caption{(c) Type 3}
        \label{fig:cervix_c}
    \end{subfigure}
\setcounter{figure}{0}
\caption{Different types of cervices based on location of the transformation zone. Type 1 is completely ectocervical, type 2 is partically ectocervical and partially endocervical and type 3 is completely endocervical.}  \label{fig:cervices}
\end{figure}

\begin{figure}
    \centering
    \includegraphics[width=0.6\textwidth]{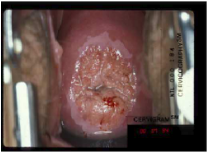}
    \caption{Specular reflections observed in the raw images of cervix~\cite{cerseg}}
    \label{fig:specular}
\end{figure}

\textbf{Challenges}: There has not been much work done in automating the detection of pre-cancerous stages of cervical cancer using raw images of cervix. This may be due to to the reason that it is challenging to capture enough features from just the raw images without having details about the patient. Also, it is very difficult to identify the type easily from the images due to specular reflections (as can be seen in Fig.~\ref{fig:specular}), blood stains, strong shady artifacts etc. The specular reflections, and other disparities in the intensity of light in the image, result from using cameras having strong flash that are used to take photographs of the cervix.

\textbf{Approach}: In this paper, we propose a method of applying the concept of deep learning and computer vision in order to automate the problem of cervical cancer screening using specular photographs of cervices. The cervical images are pre-processed by segmenting the region of interest in the cervical images using a Cervical Cancer Screening \cite{cerseg}. The images are then classified using a convolutional autoencoder based framework.  Further, due to the shortage of the number of available samples, we adopt adversarial training, by adding a discriminator to the autoencoder architecture, thus making it an adversarial autoencoder. 

The rest of the article is organized as follows - Section~\ref{Prior Art} discusses the prior art of cervical cancer screening. Section~\ref{ps} defines the problem statement that is to be solved. Section~\ref{solution} explains the details of the proposed solution. Section~\ref{obs} discusses the dataset, various experiments carried out on the dataset and their results. Section~\ref{disc} gives an explanation on the results. The conclusion is presented in Section~\ref{conc}.

\section{Prior Art} \label{Prior Art}
Several techniques for classifying cervical cancer has been proposed that use cellular features. One such technique leverage cervical biopsy that gives cellular images, from which, features were extracted and fed into a feed-forward neural network for classification~\cite{feedforward}. Multimodal features have also been used in order to perform classification. For instance, in a recent study, combined image features from the last fully connected layer of pre-trained AlexNet with biological features extracted from a Pap smear test to make the prediction~\cite{multimoddl}.  \cite{desantis} combined spectroscopic image information measured from the cervix with other patient data, such as Pap results. In \cite{coref}, an algorithmic framework based on Multimodal Entity Coreference is used for combining various tests such as Pap Test, HPV test etc. to perform disease
classification and diagnosis. Another work \cite{textdata} uses both image features and text features from various tests like Pap Smear, HPV, pH test etc., and apply separate SVMs on the two types of features in order to take a decision for the classification. 

In recent times, exploration of the deep learning paradigm has led to a surge in its use for medical diagnosis, a major advantage being that, deep learning mostly takes raw data, and we can leave the task of feature extraction to the network itself for most of the tasks. Such methods have yielded excellent results in the domain of automated cervix and cervical cell segmentation~\cite{cervixcellseg1}~\cite{cervixcellseg2}. However, there is negligible prior art which tries to classify cervical images in their pre-cancerous stages.

\section{Problem Statement} \label{ps}
The various types of cervical images during pre-cancerous stages are shown in Fig. \ref{fig:cervices}. Generally, there are three types of pre-cancerous stages of a cervix. The problem of detection of the type of cervical cancer can be therefore formally defined as a multi-class classification problem, where for each input image $I$ of size ${hxw}$, where $h$ and $w$ are the height and width of the image in pixels respectively, we have to detect the type of cervix $C$ from the cervical image, where  $C$ is one of {$0$, $1$, \dots, ${(n - 1)}$}, where $n$ refers to the total number of possible classes.
\section{Exposition to the Solution} \label{solution}
The proposed solution consists of a Convolutional Neural Network (CNN) with encoder-decoder architecture which is trained adversarially, where the encoder is not only trained to learn latent representation but also predict the class label of the input.

\begin{figure}
    \centering
    \includegraphics[width=\linewidth]{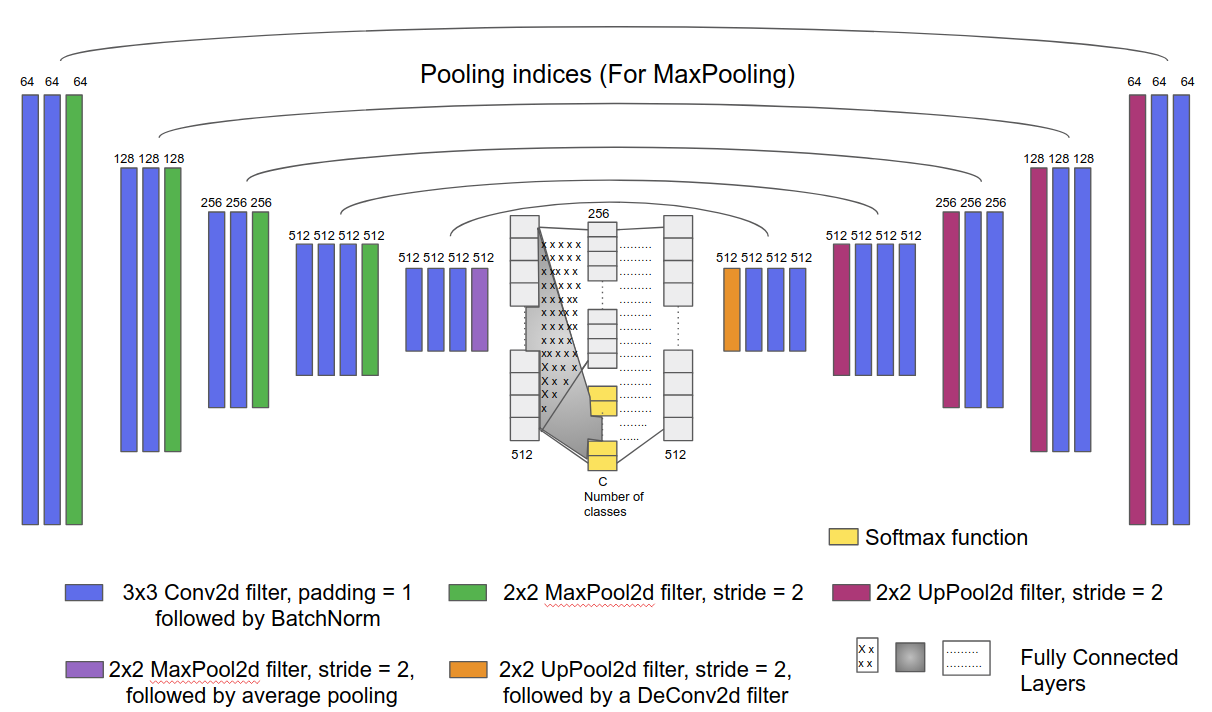}
    \caption{Feature maps of the autoencoder (The numbers on the top/bottom of the feature maps refers to the number of channels/neurons accordingly.}
    \label{fig:autoencoder}
\end{figure}

As shown in Fig.\ref{fig:autoencoder}., the proposed network is similar to that of SegNet \cite{segnet}. The encoder consists of VGG16~\cite{vgg} architecture initialized with pre-trained weights, followed by a 2D average pooling layer, which maps the tensor output of the encoder to a vector. This vector is mapped to a latent representation of length $L$ and also a categorical representation of length equal to number of classes. The categorical representation is further used to predict the class label of the image. The latent representation helps to disentangle the style of an image from its class binding \cite{adv}. The latent representation and the categorical representation vector are then concatenated and given as input to the decoder as shown in Fig.~\ref{fig:autoencoder}. Also, categorical representation vector is fed into a discriminator network along with the true class labels. The discriminator used here is just a feedforward neural network of two layers, with the input layer having number of neurons equal to the number of classes, which is fully connected to the output layer consisting of two neurons, followed by a Softmax Function\footnote{Wikipedia Page, https://en.wikipedia.org/wiki/Softmax\_function}, thus giving probabilities of the input being either real or fake as the final output.

\subsection{Training}

\begin{figure}[H]
    \centering
    \includegraphics[width=\linewidth]{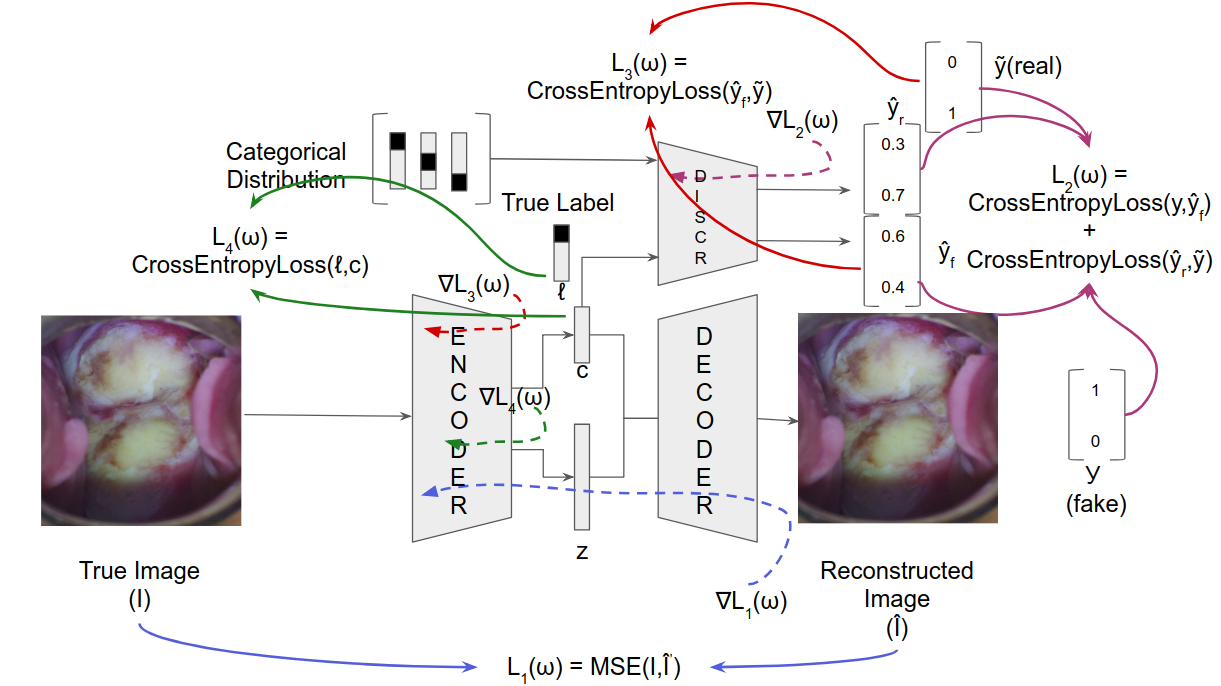}
    \caption{The figure shows the architecture of the network being used. The solid lines denote the various loss functions between corresponding inputs and targets, while, the dotted lines denote the backpropagation of gradients.}
    \label{fig:arch}
\end{figure}
Fig.\ref{fig:arch} shows the training routine. It the training consists of three phases - 
\begin{enumerate}
    \item \textbf{Reconstruction Phase}: The autoencoder produces the reconstructed image ($\mathbf{\hat{I}}$), input being the true image ($I$). The network learns by minimizing the reconstruction loss computed between the two images in this phase. The loss function used is the Mean Square Error (MSE) which is given as-
    \begin{equation}
        L_1(\omega) = \frac{\sum_{i=1}^{B} |\mathbf{\hat{I}_i} - \mathbf{I_i|^2}}{B}
    \end{equation}
    where, $\mathbf{I_i}$ and $\mathbf{\hat{I}_i}$ refers to the $i_{th}$ true image and $i_{th}$ reconstructed image respectively of the batch, $\omega$ refers to the trainable parameters of the autoencoder and $B$ refers to the batch size used for training. In the fig.~\ref{fig:arch}, $\nabla L_1$ refers to the gradients that are backpropagated through the autoencoder.
    
    \item \textbf{Regularization Phase}: In the regularization phase, firstly, the encoder output $c$ (as shown in the fig.~\ref{fig:arch}), which is the categorical representation, has to be constrained in such a way that it mimics a categorical distribution. The discriminator is trained in such a way that it can differentiate between generated/fake samples and real samples. The real samples are generated by selecting a sample from a categorical distribution, i.e., a uniform distribution of one-hot vectors having length equal to the number of classes.
    The discriminator then outputs two probabilities - one for the image being real and other for the image being fake. Let the output vector be $\mathbf{\hat{y}_r}$ represent the output vector corresponding to he real sample, and $\mathbf{\hat{y}_f}$ represent that for the fake sample. $\mathbf{\hat{y}_r}$ should ideally be $[0, 1]$ and $\mathbf{\hat{y}_f}$ should be $[1, 0]$. For this purpose, the loss function that is minimized is the sum of cross-entropy losses for the two cases, which for a single sample, can be mathematically written as -
    \begin{equation}
        L_2(\omega) = -\log(\mathbf{\hat{y}_r[1]})-\log(\mathbf{\hat{y}_f[0]})
    \end{equation}
    where $\mathbf{\hat{y}_r[1]}$ refers to the second element of the vector $\mathbf{\hat{y}_r}$ vector, $\mathbf{\hat{y}_f[0]}$ refers to the first element of the $\hat{y}_f$ and  $\omega$ refers to the trainable parameters of the discriminator. In the Fig.~\ref{fig:arch}, $\nabla L_2$ refers to the gradients that are back-propagated through the discriminator. This loss is averaged over all samples across a mini-batch.
    
    After training the discriminator, the encoder of the network is trained adversarially in order to fool the discriminator into making the output $c$ behave like a real sample, i.e., trying to map $\mathbf{\hat{y}_f}$ to $[0, 1]$. The adversarial generative loss function based on cross-entropy for a single sample, is given as,
    \begin{equation}
        L_3(\omega) = -\log(\mathbf{\hat{y}_f[1]})
    \end{equation}
    where $\mathbf{\hat{y}_f[1]}$ refers to the second element of the vector $\mathbf{\hat{y}_f}$ vector and $\omega$ refers to the trainable parameters of the encoder. $\nabla L_3$ in Fig.~\ref{fig:arch} refers to the gradients that are back-propagated through the encoder.  
    
    \item \textbf{Classification Phase}: Finally, the encoder is trained in a supervised manner, by considering only the categorical representation $\mathbf{c}$ as the output, and mapping the input to the actual one-hot label ($\ell$) corresponding to the input. The loss function optimized in this phase is the cross-entropy loss between the predicted softmax probabilities and the actual one-hot target, which is given as,
    \begin{equation}
        L_4(\omega) = -\sum_{i=1}^{C}y_ilog(S_i) 
    \end{equation}

where, $C$ is the number of classes, $S_i$ refers to the predicted probability of the input image belonging to the $i_{th}$ class, where $i \in {0,1,....,C-1}$, $y_i$ refers to the $i_{th}$ element of the one-hot target vector and $\omega$ refers to the trainable parameters of the encoder. $\nabla L_4$ in the fig.~\ref{fig:arch} denotes the gradients that are back-propagated through the encoder. 
\end{enumerate}

\section{Experiments and Results} \label{obs}

\subsection{Dataset description}
The dataset\footnote{https://www.kaggle.com/c/intel-mobileodt-cervical-cancer-screening/data} comprises of two parts - training set and test set. The training set comprises of $8,029$ specular photographs of cervix, annotated as one of the three types. The dataset is imbalanced, with $1,438$ images of type-1, $4,345$ images of type-2 and the remaining $2,426$ images of type-3. The images are of varying pixel sizes ranging from $480\times640$ to $3096\times4128$. Since test set does not have annotated labels, we have considered a subset of the training set which is held out from training for evaluation. 

\subsection{Pre-processing}
The images have different pixel sizes. Hence, all images are resized to $224\times224$. Data is then augmented by applying random rotation of upto $\pm15$ degrees on either side, and applying random horizontal flips. After this, as for the pretrained ImageNet Models, the RGB images were scaled down to pixel values in the range of $[0, 1]$ and were then normalized using mean = $[0.485, 0.456, 0.406]$ and standard deviation = $[0.229, 0.224, 0.225]$ for the red, green and blue channels respectively\footnote{https://pytorch.org/docs/stable/torchvision/models.html}. 

After this, a Cervix Segmentation Kernel is used \cite{cerseg}.
We extract the cervix region. The segmentation problem is structured as searching a contour in image, which is optimal when compared to some predefined integral measure, which is known as the the energy functional\footnote{Wikipedia Page, https://en.wikipedia.org/wiki/Energy\_functional}. The results of the cervix segmentation kernel are shown in fig.~\ref{fig:images}.
\begin{figure}[H]
\centering
    \begin{subfigure}[b]{0.3\textwidth}
        \includegraphics[width=\linewidth]{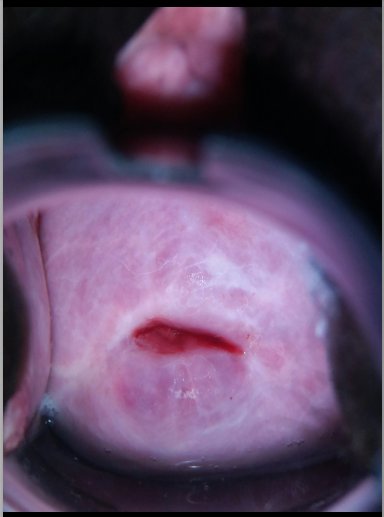}
        \caption{(a) Original Image}
        \label{fig:orig_img}
    \end{subfigure}
    \begin{subfigure}[b]{0.3\textwidth}
        \includegraphics[width=\linewidth]{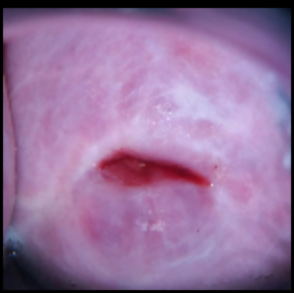}
        \caption{(b) Cropped Image}
        \label{fig:cropped_img}
    \end{subfigure}
\setcounter{figure}{4}
\caption{Using Cervix Segmentation Kernel to crop the part of the image containing cervix}
\label{fig:images}
\end{figure}

\subsection{Compensating class imbalance}
The class distribution is in the ratio $1:3.03:1.69$ i.e., the number of images of Type-2 is more than the sum of the number of images of Type-1 and Type-3. In order to balance this disparity, the loss function used in the classification phase are given class weights. Two methods for calculating the class weights were used. One is using 'balanced' class weights, where, each class weight is the reciprocal of the number of images of that type,
\begin{equation}
    CW_i = \frac{1}{n_i}
\end{equation}
where, $CW_i$ refers to the weight assigned to the $i_{th}$ class and $n_i$ refers to the number of images of the $i_{th}$ class.

The other method assigns class weights such that they are inversely proportional to the square root of the number of images of the class.
\begin{equation}
    CW_i \propto \frac{1}{\sqrt{n_i}}
\end{equation}
where $CW_i$ and $n_i$ hold the same meanings as stated in the previous case.

However, in the proposed solution, introducing class weights deteriorated the performance. This may be attributed to the fact that in the course of training, the output categorical representation of the encoder learns to mimic uniform categorical distribution, and thus, does not require class weights.

\subsection{Training Parameters}

For all the training phases, Adam optimizer is used with a learning rate of $0.0001$ and beta1 of $0.9$, beta2 of $0.999$ and a batch size of 8. The results are reported for five-fold cross validation. Training is stopped when there is no more increase in the validation accuracy.

\subsection{Baselines}

Following baselines were considered to evaluate the performance of the proposed method:

\begin{itemize}
    \item BL1 - Training the weights of only the last fully connected layer of pre-trained ResNet50 \cite{res} using class weights of second type with learning rate of 0.001, with images pre-processed using the segmentation kernel
    \item BL2 - Training the weights of all layers of pre-trained ResNet50\cite{res} with learning rate gradually decreasing from the layers at the end to that in the beginning using class weights of second type, with highest learning rate being $0.001$, and a decay rate of $0.1$ for each layer. The images were pre-processed using the segmentation kernel
    \item BL3 - Training the weights of all layers of pre-trained ResNet50\cite{res} with the same learning rate of $0.001$ using class weights of second type, without using segmentation kernel.
    \item BL4 - Training the weights of all layers of pre-trained ResNet50\cite{res} with the same learning rate of $0.001$ using class weights of second type, with images preprocessed using the segmentation kernel
\end{itemize}

\subsection{Results}
The performance of the proposed method and the baselines with respect tot accuracy, average precision and average recall are summarized in Table \ref{result_table}.

\begin{table}[H]
\centering
\caption{Performance comparison of baselines and the proposed method}
\label{result_table}
\begin{tabular}{|c|c|c|c|}
\hline
Baseline             & Accuracy & Average Precision & Average Recall \\ \hline
BL1               & 57.33\%             & 55.58\%           & 44.53\%        \\ \hline
BL2               & 57.59\%             & 54.2\%            & 49.67\%        \\ \hline
BL3               & 65.2\%              & 62.16\%           & 58.47\%        \\ \hline
BL4               & 67.72\%              & 65.87\%           & 70.25\%        \\ \hline
Proposed Solution & 73.75\%             & 75.6\%            & 73.46\%         \\ \hline
\end{tabular}
\end{table}
\section{Discussion} \label{disc}
The baselines discussed here have all been from the ResNet50\cite{res} architectures, since, ResNet50\cite{res} gave the best results when compared to other ImageNet pretrained architectures like AlexNet, VGG16, GoogleNet \cite{imagenet} etc. This may be attributed to the residual connection in ResNet50\cite{res}, which helps in the backpropagation of gradients, and hence, removes the vanishing gradient problem of deep convolutional neural networks. Also, the network was initialized with pre-trained weights. Since, the lower layers of a CNN learn very rudimentary features such as curves, edges etc. which are global in nature, it need not relearn those features. Hence, the lower layers, if initialized with pretrained weights, do not need to be changed much, which leads to faster convergence.

\begin{figure}[H]
    \centering
    \includegraphics[width=8cm]{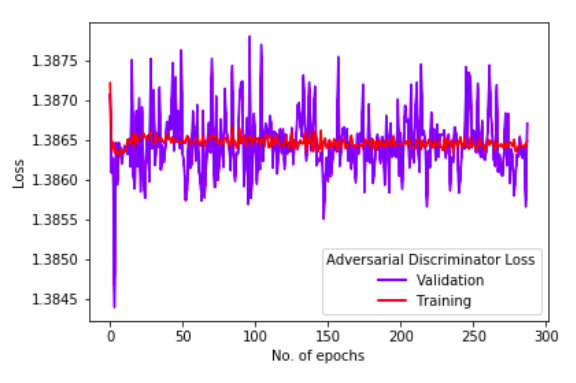}
    \caption{A plot of the discriminator loss against the number of epochs}
    \label{fig:disc}
\end{figure}

Considering the proposed solution, the plot between disciminator loss and number of epochs, as shown in in Fig.~\ref{fig:disc} suggests that, after a few epochs, the training discriminator loss reaches a nearly constant value of $1.386 \approx 2log_e{2}$, which corresponds to predicted probability of real as well as a fake input to the discriminator being 0.5. The validation discriminator loss jitters a lot initially, but eventually, its variation about the value of $2log_e{2}$ reduces, thus suggesting that with time, the discriminator is getting confused, which is desired.

\section{Conclusion} \label{conc}

\begin{figure}[H]
    \centering
    \includegraphics[width=5cm]{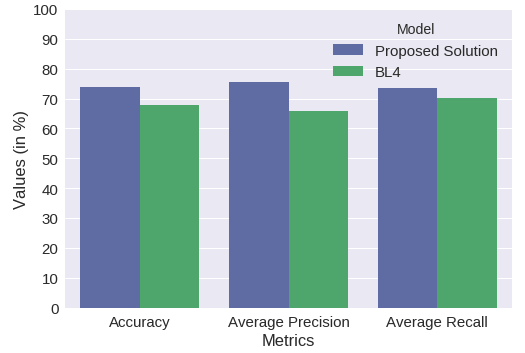}
    \caption{Comparing the results between the best baseline (BL4) and the proposed solution}
    \label{fig:metrics}
\end{figure}

The proposed solution gives a better precision and a better recall than the best baseline by more than $9\%$ and more than $3\%$ respectively, as shown in Fig.~\ref{fig:metrics}. This suggests, that our proposed solution performs better than the pretrained architectures by a good margin.

\section{Conclusion} \label{conc}

We have presented an adversarial framework for detection of the type of cervix using only the raw images of the cervix in its pre-cancerous stages. In the proposed method, an adversarially trained deep autoencoder network presented in Sec.~\ref{solution} is trained in order to perform the classification. The performance of the proposed framework is empirically verified by comparing it with some baselines, which are pretrained ImageNet architectures. It is observed that our adversarial framework outperforms the different baselines in terms of overall accuracy, average precision and average recall, giving high class-wise accuracies {$61.46\%$, $87.23\%$, $71.69\%$} for the three classes {Type-1, Type-2, Type-3}, and overall
classification accuracy of $73.75\%$.

\bibliographystyle{unsrt}  


\end{document}